# Comparative Analysis of Power System Model Reduction

Mohammad Khatibi[a], Fatemeh Rahmani[b*], Tanushree Agarwal[c]

[a] Department of Electrical Engineering, K.N.Toosi University of Technology, Tehran, Iran
[b] Department of Electronics, Central Tehran Branch, Islamic Azad University, Iran
[c] Department of Electrical Engineering, Lamar University Beaumont, TX, U.S.A

## 1. Introduction

Almost in all of the engineering applications, dynamic analysis of the system under study can be rather complex. High order and complicated mathematical equations which accurately represent the problem at hand, seems hard to be implemented practically. For example, the widespread electric power system in the world is getting more and more complicated and interconnected throughout the years especially with the new challenges introduced into power system operation and control through increasing renewable energy penetration [1, 2]. This leads to consuming a lot of time and processing memory to study the system for special purposes such as analysis, optimization or control design. Retrieving a reduced model that is not only faster in simulation, but also results in giving a correct insight into the main system is attracting attention in research and industry scopes. Power system dynamic model reduction generally consists of identifying and aggregating generators to be reduced, followed by reconfiguring the network model. Among the applied solutions in the literature, some methods stress at keeping the important eigenvalues of the main system, while others determine the reduced model in such a manner that it is, in some sense, an optimum approximation of the original system, without the constraints of matching eigenvalues or states of the main system. Many techniques and algorithms [3-6] such as Modal Analysis, coherency, slow coherency, singular value decomposition (SVD) and Gramian based methods have been proposed through the years. Also, Milner and Park introduced Bisimulation that leads to reducing the complexity of concurrent processes and studying the equivalence of automata. Eventually, the aforementioned method has been extended to include continuous-time dynamic systems [7-9].

Most technical systems are dynamic. In dynamic systems, there is no instantaneous relationship between input and output. This means that if the system's input doubles at once, for example, it will take some time for the output to reach a new value. For example, if the input voltage to a motor is increased by 20%, it will increase from 1500 rpm to 1800 rpm. Another example is that if weights are added to a mass attached to a spring, the mass is fixed after fluctuations at another point. These oscillations are called transient dynamical systems. Power systems are also usually caused by changes in frequency, voltage, real power, and reactive power after any change in input or disturbance. The main reason for most systems being dynamic is that there is an element of energy storage [10-12]. The presence of these elements makes the relationship between input and output instantaneous. For example, a capacitor in a circuit is charged by applying and removing the voltage. These examples illustrate how energy storage elements prevent the instantaneous relationship between input and output. Although energy storage elements are also of great interest, they are frustrating to a control engineer or person whose purpose is to reduce the effects of transient fluctuations. For example, if there were no energy storage elements in the power systems and the real and reactive power supply of a generator were injected to the grid, it was sufficient to change the load and voltage reference values and immediately, without any oscillations and transients, arbitrarily [13, 14].

Following a disturbance, all the synchronous machines (SMs) in an oscillating multi-machine system oscillate. Experience shows that in this case, the SMs are divided into several groups. In each group, several machines fluctuate with each other, so-called coherency behavior in machines. By checking the frequency of different machine oscillations after presence of any disturbance in the system, the same machines will behave in the same order. In this way, instead of these multi-oscillating machines, an equivalent model is inserted, which greatly reduces the specification of the one similar to that obtained in designing an overlapping model by substituting this model for system-class overlapping groups.

In the modal method, the basis is to reduce the degree of conversion of the external system into state space [15, 16]. In the system identification method, the objective is to obtain an external equivalent dynamic model, in which the basis of the discrete transformation function follows the same system. In this method, model factors are identified using methods of identification. Therefore, the degree of conversion function selection leads to a reduction of the system rank and on the other hand the important advantage of this method is that there is no need for external system factors. The only drawback is that it requires a test and recording of the information, which does not seem to be practical with the benefits of this method. Therefore, in the discussion of determining the dynamic model of the external system, this method is not preferred.

In summary, the advantages of this dynamic model reduction are as follows:

1) The noise effect and modeling error are minimized
2) Easily running with the computer
3) Not needing external system agents

Many innovative and hybrid methods have been proposed to reduce the order of dynamic systems, especially the power system. The order reduction of multi-input systems using the improved fuzzy particle mass optimization algorithm is one of the common innovative methods. One of the most important challenges in the dynamical system's hypothesis is the approximation of systems by a lower order model [17, 18].

To reduce the model order, it is easier to implement analysis, simulation, and design of different systems. Various order reduction methods are available for continuous error domain systems as well as systems in the fractional domain. Besides, single-output single-input methods have been developed to reduce the order of multi-output multi-input systems.

Different order reduction methods are based on the integral square error criterion. Nowadays, evolutionary algorithms have shown high search power which led to their promising applications to various real engineering problems [19-22] compared to previous algorithms. Additionally, these algorithms are known mainly for their ability to find an optimal solution providing us with the minimum production cost and maximum efficiency. Unlike pure mathematical methods, evolutionary algorithms do not require conditions where variables are continuous and distinctive in optimization problems. Particle mass optimization technique as an example has been used to obtain single-order single-output large-scale error systems for obtaining the order of the reduced-order model. As another example, to get the order reduction of linear dynamic systems, a particle swarm optimization beam has been studied. Modeling the order reduction of multivariate linear systems has been investigated using the PSO algorithm [23].

The present effort is to obtain an appropriate method for decreasing the order of the power system plant, while the parameters of the original power system are kept untouched and the main features of generator units are reachable. The proposed method involves searching through all the parameters to minimize the mean squares of error between the transient response characteristics of the main system and the low-order system such as ascending time, settling time.

The rest of the paper is organized as follows: in section II modal analysis is explained. In section III SVD based method is discussed. The resulted reduced-order model using these two methods explained in section IV and section V is dedicated to the conclusion.

## 2. Model Method

In this method, first, a complete model for the external system is obtained. Therefore, the input of the whole system is considered to be the voltage change and the output is considered the current change. Accordingly, the dynamic of the whole system can be written as the equations are as Eq. (1)

$$\dot{x} = Ax + B\Delta v$$
$$\Delta i = Cx + D\Delta v \quad (1)$$

In this method [24, 25], it is assumed that there exists some information about the factors of each generator. After obtaining the complete model, it is attempted to simplify the model by eliminating fast dynamics that do not reduce the accuracy of the modeling. The different stages of the procedure are as follows:

First, matrix eigenvalues and eigenvectors are obtained. Then, the transition matrix consists of Eigenvectors is formed. New state variables are defined using the resulted transition matrix. The new state-space variables for the equivalent system are obtained by performing the following calculations:

$$x = Tz \quad (2)$$

By using the transformation matrix, the resulted dynamic is as follows:

$$T\dot{z} = ATz + B\Delta v$$
$$\Delta i = CTz + D\Delta v \quad (3)$$

By simplifying the equations to get the standard state space equation:

$$\dot{z} = T^{-1}ATz + T^{-1}B\Delta v$$
$$\Delta i = CTz + D\Delta v \quad (4)$$

Which is equivalently written as following:

$$\dot{z} = A'z + B'\Delta v$$
$$\Delta i = C'z + D'\Delta v \quad (5)$$

As discussed in [26], if the matrix A doesn't have repetitive Eigenvalues, is a diagonal matrix whose diagonal elements are eigenvalues of A.

$$A' = T^{-1}AT = \begin{bmatrix} \lambda_1 & 0 & 0 \\ 0 & \cdots & 0 \\ 0 & 0 & \lambda_n \end{bmatrix} \quad (6)$$

As can be seen, the matrix is arranged in terms of dominant diameters. The eigenvalues of the system matrix are the poles of the system. The farther the poles on the plane, the smaller the axis will have less effect on the dynamical behavior of the system. In this section, we can avoid the effect of several poles, which are very far from the imaginary axis, by first dividing the eigenvalues into two important and economical parts.

$$A'' = \begin{bmatrix} \lambda_1' & 0 & 0 \\ 0 & \cdots & 0 \\ 0 & 0 & \lambda_n' \end{bmatrix} \quad |\lambda_1'| \leq |\lambda_2'| \leq \ldots \leq |\lambda_n'|$$

$$\Delta i = \begin{bmatrix} C_1'' & C_2'' \end{bmatrix} \begin{bmatrix} W_1 \\ W_2 \end{bmatrix} + D''\Delta v \quad (7)$$

And consequently, we have:

$$\dot{W}_1 = A_1''W_1 + B_1''\Delta v$$
$$\dot{W}_2 = A_2''W_2 + B_2''\Delta v$$
$$\Delta i = C_1''W_1 + C_2''W_2 + D''\Delta v \quad (8)$$

So far, no approximations have been made in the calculations, only the shape of the model has changed. Then the reduced system is as follows:

$$\dot{W}_2 = 0 \Rightarrow W_2 \equiv -A_2''^{-1}B_2''\Delta v$$
$$\dot{W}_1 = A_1''W_1 + B_1''\Delta v$$
$$\Delta i = C_1''W_1 + (D'' - A_2''^{-1}B_2'')\Delta v \quad (9)$$

This method is suitable for dynamic stability studies but is not usually suitable for transient stability studies. One drawback of this method is the difficulty of obtaining the relation. Another disadvantage of this method is the need to know the specifics of each component of the system [27, 28].

## 3. SVD Method

In this section, according to an SVD based model reduction algorithm is introduced that produces a reduced model S by projection with the matrix Z having the specific form: $Z = QV(V^TQV)^{-1}$. Where Q is the observability Gramian as defined in (2.1) and V spans a rational Krylov subspace. The specific choice of V will be explained below.

Clearly, $Z^T V = I_r$ and the reduced systems in state-space is given by:

$$S_r := \begin{bmatrix} A_r & b_r \\ C_r & 0 \end{bmatrix} = \begin{bmatrix} (V^TQV)^{-1}V^TQAV & (V^TQV)^{-1}V^TQb \\ CV & 0 \end{bmatrix} \quad (10)$$

In the reduction step, V reflects the Krylov-side of the algorithm and Z reflects the SVD-side. With the choice of Z, the quality of the approximant $S_r$ critically depends on the reducing subspace V; consequently, the interpolation points used to form V. In this note, in constructing the rational Krylov subspace V, we will choose the interpolation points in a (sub) optimal way based on the following theorem, a straightforward extension of Gaier's discrete-time result to multivariate continuous-time systems. The SVD-rational Krylov based model reduction method algorithm is as follows:

I. Make an initial shift selection, for i= 1. . . r.
II. Construct V such that:

$$R_{an}(V) = \text{Span}\left(\left[(s_1I_n - A)^{-1}g, \ldots, (s_rI_n - A)^{-1}b\right]\right)$$

with $V^TV = I_r$

III. $Z = QV(V^TQV)^{-1}$

while (the relative change in $s_i \geq tol$)

IV. (a) $A_r = Z^TAV$

V. (b) $s_i \leftarrow -\lambda_i(A_r)$ for i=1,...,r

It follows that upon convergence, $s_i = -\lambda_i(A_r)$, for $i = 1..r$; and hence $G_r(s)$ interpolates $G(s)$ at the mirror images of the reduced poles, as desired. We note that the orthogonalization of V in Steps 2 and 4(c) above are for numerical purposes only. Instead, one can simply set

$$V = \left[(s_1I_n - A)^{-1}b, \ldots, (s_rI_n - A)^{-1}b\right].$$

## 4. Simulation Results

The proposed linearization, the modal truncation, and SVD based algorithms to identify whether the power system can maintain the synchronism after a large disturbance is applied to the well-known New England 10-generator, 39-bus benchmark system. Input data of the New England test system are provided in [29, 30]. A schematic of the test system is shown in Figure 1. Three-phase faults at buses that occurs at 1 s and cleared at 1.1 s (5 cycles) are applied as a disturbance to the system. The duration of the study period is 20 s, the time step length for the numerical integration is set to $\Delta t = 0.005$. Figure 2. Illustrate a comparative overview of the value of the rotor angle of SM $G_1$ for specified disturbance when using the original nonlinear model and the reduced linearized dynamic models. The initial values of the order of the linearized model were set to $r = 60$.

Table 1. Shows a comparative overview of the eigenvalues of the original and reduced model, after linearizing the nonlinear model of the power network and deriving the state-space matrix.

Because of the fast decay of eigenvalues that have real parts far from the origin, they are not listed in the table. Also, repeated eigenvalues are considered just once. However, eigenvalues are not being explicitly retained, it is interesting to see whether the slowest modes are kept in the reduced model. In fact, from Table 2. It can be inferred that the two slowest modes are reproduced precisely in the reduced model, with the second and third slowest modes being strongly overestimated in terms of their damping. In this section, the accuracy of the aforementioned methods has been evaluated in both the time and frequency domain.

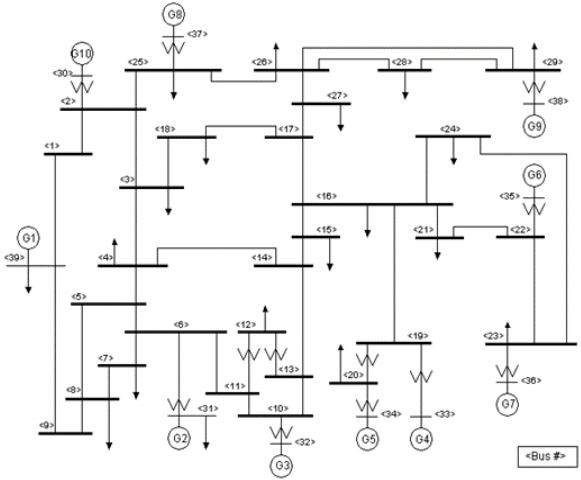

**Figure 1.** New England 10 generator 39 bus benchmark

The proposed linearization algorithm to identify whether the power system can maintain the synchronism after a large disturbance is applied to the well-known New England 10-generator, 39-bus benchmark system. Input data of the New England test system are provided in [31, 32]. Figure 1. shows the schematic of the test system.

**Table 1.** The main charactristics of the dynamic model for the new england system

| SMs | Automatic Voltage regulators | Turbine-governors | Buses | Total |
|---|---|---|---|---|
| *Number of units* | | | | |
| 1 | 9 | 10 | 10 | 39 |
| *Number of state/algebraic variables* | | | | |
| 3/3[a] | 4/3[b] | 4/1[c] | 3/2[d] | 0/2 | 109/138 |

The main characteristics of the dynamic model are summarized in Table 1. three-phase faults at buses 3 occurring that occur at 1 s and cleared at 1.1 s (5 cycles) is applied as a disturbance to the system. The duration of the study period is 15 s, the time step length for the numerical integration is set to $\Delta t = 0.005$.

Figure 2. shows the rotor angle comparison for generator 1 for the original and reduced model using the modal reduction method.

| Full system | Reduced System | Full System | Reduced System | Frequency Error (%) |
|---|---|---|---|---|
| 0.0035 | 0.0035 | -0.7726 | -0.7727 | 0 |
| 0.0103 | 0.0103 | -0.7321 | -0.7321 | 0 |
| 0.0266 | 0.0262 | -0.587 | -0.5896 | 1.5038 |
| 0.0335 | 0.0337 | -0.5499 | -0.5499 | -0.5970 |
| 0.0678 | 0.0678 | -0.5275 | -0.5275 | 0 |
| 0.1981 | 0.1981 | -0.4848 | -0.4846 | 0 |
| 0.2328 | 0.2328 | -0.4315 | -0.4315 | 0 |
| 0.5636 | 0.5636 | -0.4141 | -0.4141 | 0 |
| 0.8424 | 0.8424 | -0.398 | -0.3981 | 0 |
| 0.9162 | 0.9162 | -0.3859 | -0.3859 | 0 |
| 0.9379 | 1.0234 | -0.3587 | -0.3587 | -9.1161 |
| 1.0234 | 1.0449 | -0.3272 | -0.3272 | -2.1008 |
| 1.0449 | 1.1118 | -0.2703 | -0.2704 | -6.4025 |
| 1.1118 | 1.3238 | -0.262 | -0.2621 | -19.0682 |
| 1.3237 | 1.3606 | -0.232 | -0.2319 | -2.7876 |

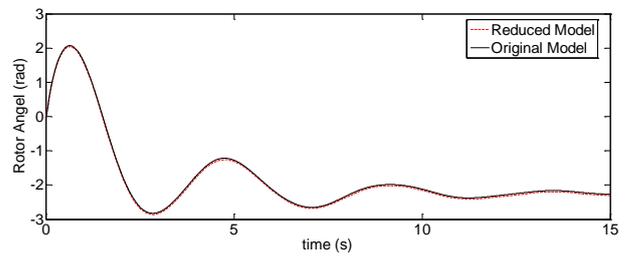

**Figure 2.** Rotor angle comparison for SMG1 between original and reduced model

**Table 2.** Modes of the original and the reduced systems

Figure 3. shows the rotor angle comparison for generator 1 for the original and reduced model using the SVD reduction method.

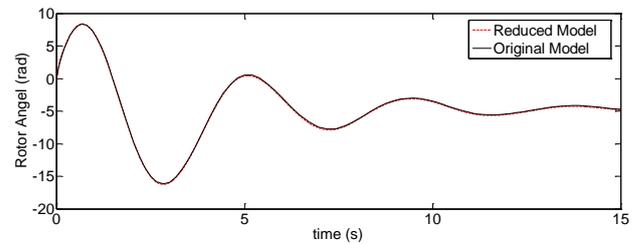

**Figure 3.** Rotor angle comparison for SMG1 between original and reduced model

Figure 4. shows the comparison for the first diagonal term of generator 1 for the original and reduced model using the modal reduction method.

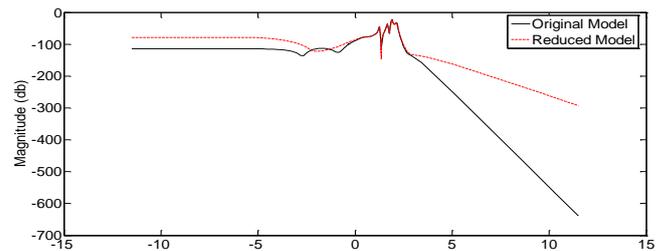

**Figure 4.** 1st diagonal term of transfer function

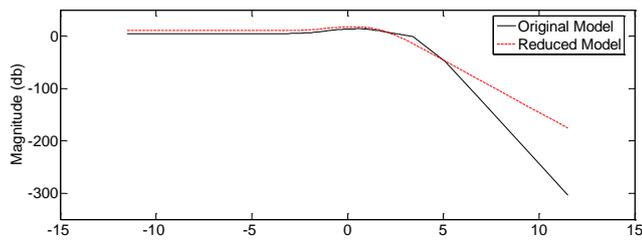

**Figure 5.** 1st diagonal term of transfer function

Figure 5. shows the comparison for the first diagonal term of generator 1 for the original and reduced model using the SVD reduction method.

## 5. Conclusions

Dynamic model reduction of the power system as any complicated system has many advantages such as resulting in less computational burden and fast simulation that leads in easier analysis and control. Among extensive available methods in the literature that are mainly discussed in control theory and mathematics, modal analysis and SVD were briefly reviewed in this paper. To validate the results, the simulation is performed on the New England test system and the stability of the reduced model is guaranteed. Modal dominance analysis, time-domain behavior of generator rotor angle and frequency response during and after a disturbance were hired to check the validity of the method. Therefore, without performing the potentially expensive eigenvalue decomposition simulation, this method seems to adequately reproduce the dominant modes of the original system guided by their influence in its input-output behavior. Preserving the coherency between generators in each coherent group is also shown after the reduction method is applied.